\documentclass[24pt]{article}
\usepackage{amssymb}
\usepackage{epsfig,caption,graphicx,eepic,epic}
\usepackage{pstricks}
\usepackage{amssymb,amsmath}
\usepackage{multido}
\usepackage{array}
\begin{document}
%\draft
\def\be{\begin{equation}}
\def\ee{\end{equation}}
\def\ba{\begin{eqnarray}} 
\def\ea{\end{eqnarray}}
\def\nn{\nonumber}
\newcommand{\bbf}{\mathbf}   
\newcommand{\rrm}{\mathrm}
\title{\bf A short note about the dynamical description of the measurement process in quantum physics}
\author{Jean Richert$^{a}$
\footnote{E-mail address: j.mc.richert@gmail.com}\\ 
and\\
Tarek Khalil$^{b}$
\footnote{E-mail address: tkhalil@ul.edu.lb}\\ 
$^{a}$ Institut de Physique, Universit\'e de Strasbourg,\\
3, rue de l'Universit\'e, 67084 Strasbourg Cedex,\\      
France\\
$^{b}$ Department of Physics, Faculty of Sciences(V),\\
Lebanese University, Nabatieh,
Lebanon} 
 
\date{\today}
\maketitle 
\begin{abstract}
The measurement process of observables in a quantum system comes out to be an unsovable problem which started in the early times of the development of the theory. In the present note we consider the measured system part of an open system interacting with the measuring device and show under which ideal conditions the measure process may ideally work. Our procedure leads to the conclusion that there is no hope that any experimental procedure will be able to lead to a clean solution of the process, except maybe in very specific cases. The reasons for this situation are deeply rooted in the fundamental properties of quantum theory.
\end{abstract}


\section{Introduction}
In a measurement process of observables a quantum system is in contact with a physical environment, in practice a measuring device which is aimed to report the quantitative value of the microscopic observable of the system under scrutiny. It comes out that the measurement of a physical observable by means of an experimental device leads to fundamental problems which have been considered already by the fathers of the quantum theory for no less than a whole century and never satisfactorily solved for reasons which are deeply rooted in the properties of the theory. As a consequence a huge controversial litterature developed on the subject starting already at the very beginnigs of the development of the quantum theory about a century ago ~\cite{wh,nb}.\\

A clean formulation of a measuring process can be essentially prsented in the framework of an open quantum system coupled to an environment, an instrument, which is aimed to provide the quantitative determination of the physical observables representing the measured quantities. The aim of the present work consists in the determination of the conditions under which such an ideal measurement is in principle possible. The development of the formalism will show under what conditions this goal could in principle be reached and, at the same time, lead to the well-known result that it will generally never be the case.\\

\section{Mathematical formulation of the measuring process}

In the quantum formalism physical quantities are represented by operators called observables, say $A$ and $B$, which do not necessarily commute

\ba
[A,B] \neq 0
\label{1}
\ea

This mathematical property generates the concept of incompatibility between $A$ and $B$. Contextuality results from the fact that in this case it is not possible to assign simultaneously a single joint probability to both observables $A$ and $B$. This fact lies at the roots of quantum theory and has been multiply examined ~\cite{gr,bu,wo,se,ca,gl,sc,kh,or,la,er,yu}.\\

Among many attempts to formulate the measure problem a quantum formalism which fixes the mathematical framework of a measure of the theory has been worked out in the $80's$~\cite{ma} and extended some time later~\cite{ka,yui}. This formalism did however not explicitly insist on the inherent difficulties which enter the problem except for the repeatability problem, the possibility to reproduce sequentially the measure of the value of a given physical observable. In the following we shall now show the strong constraints under which this problem can find a rigorous theoretical solution. 

\subsection{Evolution of a system coupled to an environment}
We consider a system $S$ governed by a Hamiltonian $H_{S}$ coupled to an external quantum device $M$ governed by a Hamiltonian $H_{M}$, in dynamic interaction with $S$ by means of a coupling operator $H_{C}$.
The Hamiltonian of the whole system reads

\ba
H=H_{S}+H_{M}+H_{C}
\label{2}
\ea

The time evolution of the whole system in the Hilbert space of $S + M$ is governed by the density operator $\omega(t)$, at t=0 $\omega(0)=\rho_{S}(0) \otimes \mu_{M}(0)$  where $\rho_{S}(t)$ and  $\mu_{M}(t)$ act in $S$ and $M$ spaces. The general expression of the evolution of $\omega(t)$ which describes the evolution of the total system and takes care explicitly of the coupling of the system to the maeasure device is given by

\ba
i\frac{d\omega(t)^{ik}_{\lambda \nu}}{dt}=\sum_{j}[H_{S}^{ij}\otimes I_{\lambda \nu},\omega^{jk}_{\lambda \nu}]          
+\sum_{j \mu}[H^{ij}_{C \lambda \mu},\omega^{jk}_{\mu \nu}]+\sum_{\mu}[I_{ik}\otimes H_{M}^{\lambda \mu},
\omega^{ik}_{\mu \nu}]                    
\label{3}
\ea
where the elements  $I_{ik}$ and $I_{\lambda \nu}$ are the components of the unity operators $I_{S}$ and $I_{M}$ in $S$ and $M$ spaces respectively. The observable which has to be measured lives in $S$ space, the measurement outcome is generated in $M$ space. 

\section{Measurement process of quantum observables}
 
and\\
The time evolution of the interacting subsystems $S$ and $M$ is put here in its most general form. In the present work subsystem $M$ is now understood as the quantum system which is devoted to the measure of a physical observable $O$ defined in $S$ space and measured in $M$ space.\\

Let $O$ be an observable defined in a complete basis of states ($|i\rangle, i=1,N$) of $S$ space. The observables defined in this space are  measured with an apparatus which works in $M$ space. The question is whether this can be done rigorously and under what conditions. As it is known this is generally not possible, because of the specific properties of the quantum theory which introduces the non commutability of physical operators (see the prolific literature below).There exists however a way out of this situation if one is able to realize rhe experiment under strong formal conditions.\

Going back to Eq.{3} we consider the case where the system $S$ is prepared in a diagonal basis of states at $t=0$ and ask for the commutativity of $H_{S}$ and $H_{C}$

\ba
[H_{S} \otimes I_{M}, H_{C}]=0
\label{4}
\ea

As a consequence of this condition the physical system described by $\omega(t)$ will stay in the state 
$|i\rangle$ in which it was generated at $t=0$ ~\cite{kr1,kr2}.

Similarly one may ask that the measure system $M$ is prepared in a diagonal basis of states at $t=0$ and 

\ba
[H_{C}, H_{M}\otimes I_{S}]=0
\label{5}
\ea

As a consequence the physical quantity $\langle i|O|i \rangle = O_{i}$ corresponding to the observable $O$ will stay in state $|i \rangle$ when time flows.

Under these conditions a rigorous measure of the observable $O$ in state $|i\rangle$ defined in $S$ space can be done in $M$ space by the apparatus. Indeed, the pure state $|O_{i}\rangle \otimes |m_{\lambda}\rangle$ defined in ${\cal H}_S \otimes {\cal H}_{M}$ at $t=0$ staying constant over any interval of time $[0,\tau]$ the pointer of the device prepared for the measure of the observable $O_{i}$ will deliver a result, say a real quantity, 
$c_{i \lambda}$ and the measure operation can be repeated for any further time $\tau+\Delta \tau$.
If however Eqs.[4] and [5] are not verified different measure processes generally lead to different answers of the measure pointer.The source of this difficulty is due to the fact that in that case each measurement procedure may lead to a different pointer state and a sharp way measure value will be replaced by a distribution of measurement values. This is explicitly seen by looking at expressions (36) and (37) of paragraph 5.5 in~\cite{kr2}, see also~\cite{kr1}.

Going through different runs the accumulation of the results may be collected and by repetition one may get a distribution of pointer results leading to a statistical distribution  
\ba
\sigma_{i}=\sum_{\lambda} p_{\lambda}c_{i \lambda}
\label{6}
\ea
where the summation is done over a number of measures and the $p_{\lambda}$ are the statistical weights of the resulting distribution. Recent approaches of the question concerning measurement incompatibilty can be found in refs.~\cite{nin,ku}.

\section{Discussion}

The preceding considerations exemplify the fundamental difficulty of a measurement in the quantum framework. Clearly the derivation which results from Eqs.(4-5) corresponds to the exceptional case where the involved operators commute. This is however quite never the case.

Starting from the very beginnings of the quantum description it became clear that the development of the theory ~\cite{wh,nb} leads to serious difficulties related to specific properties of the mathematical formalism implied, i.e. properties like the non-commutativity of observables which hinders the determination of precise values of physical quantities in terms of the corresponding observables.

This situation opened the way to a considerable amount of developments and interpretations relying on the definition of theoretical concepts such as incompatibility, and contextuality ~\cite{kh,gj} which lead to the impossibility to measure a physical quantity in a single step and forbids the repetition of the process 
ref.~\cite{vn,ga}. A huge amount of theoretical work has been proposed in recent times in order to overcome this still open and seemingly rigorously unsolvable problem (see the recent bibliography below).\\  
We refer in particular to the work developed by F.A. Mueller, see ref.~\cite{mu}, more precisely chapter 5 for a detailed development of some of these approaches, the work of Govind Krishnan ~\cite{gkr} and 
Laszlo E.Szabo and coll.~\cite{sgg}.\\

\end{document}